\documentclass[11pt]{article}

\usepackage{fullpage}
\usepackage{times}
\usepackage{amsmath,amssymb,amsfonts,amsthm} 
\usepackage{graphicx}
\usepackage{xspace}

\newtheorem{definition}{Definition}
\newtheorem{proposition}[definition]{Proposition}
\newtheorem{lemma}[definition]{Lemma}
\newtheorem{theorem}[definition]{Theorem}

\newtheorem{claim}[definition]{Claim}
\newtheorem{example}[definition]{Example}
\newcommand{\email}[1]{{\tt #1}}

\newcommand{\ctr}{\mathrm{ctr}}

\newcommand{\StableMatch}{{\sc StableMatch}\xspace}

\title{General Auction Mechanism for Search Advertising}

\author{
Gagan Aggarwal\thanks{Google, Inc., 1600 Amphitheatre Pkwy, Mountain View, CA, 94043. \email{gagana@google.com}}
\and
S. Muthukrishnan\thanks{Google, Inc., 76 Ninth Avenue, 4th Floor, New York, NY, 10011. \email{muthu@google.com}}
\and
D\'avid P\'al\thanks{David R. Cheriton School of Computer Science, University of Waterloo, Waterloo, ON, Canada. \email{dpal@cs.uwaterloo.ca}.
Work done during summer 2007 internship at Google New York.}
\and
Martin P\'al\thanks{Google, Inc., 76 Ninth Avenue, 4th Floor, New York, NY, 10011. \email{mpal@google.com}}
}

\date{July 3, 2008}

\begin{document}

\maketitle


\begin{abstract} 
In sponsored search, a number of advertising slots is available on a
search results page, and have to be allocated among a set of
advertisers competing to display an ad on the page. This gives rise 
to a bipartite matching market that is typically cleared 
by the way of an automated auction. Several
auction mechanisms have been proposed, with variants of the
Generalized Second Price (GSP) being widely used in practice.

There is a rich body of work on bipartite matching 
markets that builds upon the stable marriage model of Gale and Shapley
and the assignment model of Shapley and Shubik. 
This line of research offers deep insights into the structure of stable 
outcomes in such markets and their incentive properties.

In this paper, we model advertising auctions in terms of an
assignment model with linear utilities, extended with bidder and item
specific maximum and minimum prices. Auction
mechanisms like the commonly used GSP or the well-known Vickrey-Clarke-Groves (VCG) can be interpreted
as simply computing a \emph{bidder-optimal
stable matching} in this model, for a suitably defined set of bidder
preferences, but our model includes much richer bidders and preferences. 
We prove that in our model the existence of a stable
matching is guaranteed, and under a non-degeneracy assumption a
bidder-optimal stable matching exists as well. We give a fast algorithm to
find such matching in polynomial time, and use it to design truthful
mechanism that generalizes GSP, is truthful for profit-maximizing
bidders, correctly implements features like bidder-specific minimum
prices and position-specific bids, and works for rich mixtures of bidders and preferences. 
Our main technical contributions
are the existence of bidder-optimal matchings and (group)
strategyproofness of the resulting mechanism, and are proved by induction
on the progress of the matching algorithm.

\end{abstract}

\section{Introduction}
\label{section:introduction}

Internet advertising is a prime example of a matching market: a number 
$n$ of advertisers (bidders) are competing for a set of $k$ 
advertising slots (items) offered for sale by a content publisher or a 
search engine. Internet advertising and sponsored search auctions have 
attracted wide attention in the academic literature, 
and there are several papers discussing various aspects of pricing 
ad slots and allocating them to interested advertisers.

Classical matching market models include the 
\emph{stable marriage} model of Gale and Shapley \cite{GaleShapley1962}
and the \emph{assignment model} of Shalpley and Shubik \cite{ShapleyShubik1971}.
For these models and many of their extensions, we have a good 
understanding of the structure of their stable outcomes (``equilibria'')
and their incentive properties. We take advantage of existing body
of work on stable matchings and apply it to sponsored search.

We observe that existing auction mechanisms for sponsored search, most notably, variants of Generalized Second Price (GSP) and Vickrey-Clarke-Groves (VCG), merely compute a stable 
matching in a suitably defined model. We make this model explicit, and
propose a new auction mechanism that includes the existing mechanisms as 
special cases. The model is flexible enough to allow for
bidder and position specific minimum and maximum prices, as well as 
different values for different slots. Much of the existing literature does not address these 
features (like minimum prices) that are important in practice.
Beyond that, our model of bidder preferences allows for a wider range of bidder behaviors
than just  profit maximization (i.e. we do not assume that the bidder's payoff
is quasi-linear in payment). As an example, a bidder who desires to win 
the highest slot possible subject to the constraint that his price be 
at most $m$ (for some parameter $m$) is clearly not maximizing profit, but
can be expressed in our model. It is important for us to include such bidders
in order to correctly model the variants of GSP auction which have not been previously analyzed; it 
also happens that the basic GSP mechanism is truthful for such class of bidders. 

Our proposed auction mechanism solicits bidder preferences from each bidder 
and then simply computes a
\emph{bidder-optimal stable matching} given those preferences.
The mechanism is truthful (and even group strategyproof if money transfers
among players are not permitted). 

On the algorithmic side, we show how to compute the allocation and
prices corresponding to a bidder-optimal stable matching in time
$O(nk^3)$, where $n$ is the number of bidders and $k$ is the number of
slots to sell. Our algorithm is an extension of the Hungarian
algorithm for finding maximum-weight matchings in bipartite
graphs. The idea of the algorithm is simple, although some attention to
detail is required to ensure correctness, and the algorithm has to be made fast enough
for search advertising.
Our proofs of existence of
bidder-optimal matchings as well as proof of a key lemma establishing
truthfulness of our auction mechanism follow by induction on the
execution of the matching algorithm.

\section{Related Work}
\label{section:related}

\paragraph{Matching Markets.}
The marriage model of Gale and Shapley \cite{GaleShapley1962} 
and the assignment model of Shapley and Shubik
\cite{ShapleyShubik1971} are two standard models in the theory of 
matching markets. 

In the marriage model, a set $I$ of men and a set $J$ of women
is given, where each man and woman is endowed with a ranked list of
members of the opposite sex.  Men and women are to be matched in a one
to one fasion. A matching is considered stable if there is no man and
a woman who would simultaneously prefer each other to their respective
assigned partners. A stable matching is guaranteed to exist, and 
the \emph{deferred acceptance} algorithm can be used to find it.
The stable matching found by this algorithm is \emph{man-optimal}, in that 
every man prefers it to any other stable matching.
Moreover when using the deferred acceptance algorithm, no man has 
an incentive to misreport his true preference order 
\cite{Roth1982}.

The assignment model \cite{ShapleyShubik1971}, (see also
\cite{Quinzii1984, DemangeGaleSotomayor1986})
differs in that each player derives a certain value
from being matched to each person of the opposite sex, and side payments
between partners are allowed. The goal of each player is to maximize his or 
her payoff which is the sum of partner's value and monetary payment
(positive or negative negative) from the partner.
The set of stable outcomes is non-empty by a linear programming
argument. In fact, each stable outcome corresponds to a maximum-weight
matching, and player payoffs correpond to dual variables of the maximum 
matching LP. A man-optimal outcome is guaranteed to exist, and its
allocation and prices are identical to the VCG mechanism
for maximum weight matchings \cite{Leonard1983,BikhandaniOstroy2006}.

Many variations and extensions of each model have been studied; 
see the monograph \cite{RothSotomayor1990} for a nice overview.
Payoff functions that are not necessarily linear in the payment were
considered by~\cite{DemangeGale1985, DemangeGaleSotomayor1986,
Alkan1989, AlkanGale1990}. 
Even in such generality,
there exists a man-optimal stable matching~\cite{DemangeGale1985}, 
and in a man-optimal auction mechanism, it is weakly dominant for each bidder 
to reveal his true utility (payoff) function.
These results require the utility functions to be continuous, strictly monotone 
and defined on the whole range $(-\infty, +\infty)$, and therefore are not directly 
applicable in our setting.

Kelso and Crawford \cite{KelsoCrawford1982} and others have proposed a many 
to one variant in which firms may hire multiple workers. 
Recently, Fujishige and Tamura \cite{FujishigeTamura2007} proposed a 
very general many to many
model with linear utility functions in which each worker
can engage multiple firms, and allow lower and upper bounds to be placed
on the range of payments allowed between any pair of players.
Under an assumption on the  payoff functions called $M^\natural$ concavity,
they give a proof of existence of a stable outcome and give an algorithm 
to find it.

The model considered in this paper
is an assignment model with linear payoffs. It is a special 
case of the model of Fujishige and Tamura \cite{FujishigeTamura2007},
in that we assume one to one matching of bidders to items.
In addition to non-emptiness, we show that the set of stable matchings in our
model has a bidder-optimal element, and prove that an auction mechanism 
based on bidder-optimal matchings is truthful and present an efficient algorithm. 
Fujishige and Tamura \cite{FujishigeTamura2007} 
show existence of a stable matching in their very general model 
by running an algorithm somewhat similar to ours, 
but do not give any results on bidder-optimality or truthfulness.

\paragraph{Sponsored Search Auctions.}
Flavors of the Generalized Second Price (GSP)
auction are the dominant vehicles for selling ads on the 
internet. In its basic form, GSP solicits a numeric bid from each advertiser, 
orders them in decreasing order of bids, and assigns slots to the first up to
$k$ bidders in this order. Each bidder is required to pay a price 
equal to the bid of the next bidder in the ordering (or a minimum price
if this is the last bidder). In a per-click GSP, each bidder pays only in the 
event that his ad is clicked on. In a per-impression GSP, the advertiser pays
each time her ad is displayed.

It has been observed that although it is not truthful for ``profit
maximizing'' bidders, the per-click GSP mechanism does have 
a Nash equilibrium (under some assumptions on the structure of click 
probabilities across different positions) that is efficient and its 
resulting prices are equal to VCG prices; see
\cite{EdelmanOstrovskySchwarz2007,AggarwalGoelMotwani2006}. A
variant of GSP in which the bidder can specify the lowest (maximum)
acceptable position has been proposed in
\cite{AggarwalMuthukrisnanFeldman2006}, which also has a Nash
equilibrium equivalent to a suitably defined VCG auction.  
Even-Dar et al.~\cite{EvenDarMansour2008} show that a Nash equilibrium 
of GSP exists even if minimum prices are bidder-specific, but that equilibrium
is no longer related to a naturally defined VCG outcome.

One reason GSP works well in practice is that in most situations, bidders
universally agree that higher slots are preferable to lower slots.
With increasingly complex web page layouts and increasingly sophisticated 
advertisers this assumption may become less valid over time. Features 
like Google's Position Preference aim to rectify this by allowing advertisers
to only bid for a specified subset (range) of positions.

The general class of VCG mechanisms follows from 
works of Vickrey~\cite{Vickrey1961}, Clarke~\cite{Clarke1971} and 
Groves~\cite{Groves1973}. For an overview of 
the VCG mechanism applied to sponsored search, see e.g. 
\cite{Aggarwal2005,AggarwalGoelMotwani2006}. VCG is a very natural
mechanism and is truthful for profit maximizing bidders, but it is 
sufficiently different from GSP and bidders may find it difficult 
to interpret the prices they are charged. 

In section \ref{section:model} we describe the assignment model with minimum 
and maximum prices and state the main results. 
Section \ref{section:algorithm} gives a description of an algorithm
to find a bidder-optimal stable matching. Sections 
\ref{section:bidder-optimality} and \ref{section:truthfulness}
give high level overview of the proofs, with the details delegated to 
Appendix \ref{section:bidder-optimality-proofs} and 
\ref{section:incentive-compatibility-proofs}. 
Appendix \ref{section:new-auctions} discusses how current auction mechanisms
for sonsored search fit in our model.

\section{Assignment Model with Maximum and Minimum Prices}
\label{section:model}

Our model that we call the {\em max-value} model, consists of the set $I=\{1,2,\dots,n\}$ of bidders and the set
$J=\{1,2,\dots,k\}$ of items. We use letter $i$ to denote a bidder and 
letter $j$ to denote an item.
Each bidder $i$ has a \emph{value} $v_{i,j}$ for each
slot $j$ how much is that slot worth to her, and a \emph{maximum
price} $m_{i,j}$ she is able and willing to pay for the slot.\footnote{To
motivate why $v_{i,j}$ and $m_{i,j}$ might be different, consider buying
a house whose value to you is higher than the amount of money your bank
is willing to lend you. Allowing the bidder to specify
both a value and a maximum is also needed to model the GSP auction.}
In addition to bidder preferences, the seller may specify for each item $j$ 
a \emph{reserve} or \emph{minimum price} $r_{i,j}$.  

For simplicity we assume that the minimum prices are known to the
bidders in advance.  For each $i$ and each $j$ we assume that $r_{i,j}
\ge 0$, $v_{i,j} \ge 0$, $m_{i,j} \le v_{i,j}$.  If bidder $i$ is
interested in the slot $j$ he specifies $m_{i,j} \ge r_{i,j}$.
Otherwise, if bidder $i$ has no interest in slot $j$ he specifies
negative $m_{i,j}$.  We denote by $v,m,r$ the $n \times k$ matrices
with entries $v_{i,j}, m_{i,j}, r_{i,j}$ respectively. We refer to the
triple $(v,m,r)$ as an \emph{auction instance} or simply
\emph{auction}.

\paragraph{Stable Matching.}
We formalize the notion of a matching in the following definitions.

\begin{definition}[Matching]
\label{definition:matching}
A \emph{matching} is a triple $(u,p,\mu)$, where
$u=(u_1, u_2, \dots, u_n)$ is a non-negative \emph{utility vector},
$p=(p_1, p_2, \dots, p_k)$ is a non-negative \emph{price vector}, and 
$\mu \subseteq I \times J$ is a set of bidder-slot pairs such that
no slot and no bidder occurs in more than one pair.
\end{definition}

If a pair $(i,j)\in \mu$, we say that bidder $i$ is \emph{matched} to slot
$j$. We use $\mu(i)$ to denote the slot matched to a bidder $i$,
and $\mu(j)$ to denote to denote the bidder matched to a slot $j$.
Bidders $i$ and slots $j$ that do not belong to any pair in $\mu$
are said to be \emph{unmatched}.

\begin{definition}[Feasible matching]
\label{def:feasibility}
A matching $(u,p,\mu)$ is said to be \emph{feasible} for an auction 
$(v,m,r)$, 
whenever for every $(i,j) \in \mu$, 
\begin{align}
\label{equation:feasibility1}
p_j \in [r_{i,j}, m_{i,j}] \; , \\
\label{equation:feasibility2}
u_i + p_j = v_{i,j} \; ,
\end{align}
and for each unmatched bidder $i$ is $u_i = 0$ and for each unmatched
slot $j$ is $p_j = 0$.
\end{definition}

\begin{definition}[Stable matching]
\label{def:stability}
A matching $(u,p,\mu)$ is \emph{stable} for an auction $(v,m,r)$
whenever for each $(i,j) \in I \times J$ at least one of the following
inequalities holds:
\begin{align}
\label{equation:stability1}
u_i + p_j &\ge v_{i,j} \; , \\
\label{equation:stability2}
      p_j &\ge m_{i,j} \; , \\
\label{equation:stability3}
u_i + r_{i,j} &\ge v_{i,j} \; .
\end{align}
A pair $(i,j) \in I \times J$ which does not satisfy any of the three
inequalities is called \emph{blocking}.
\end{definition}
\begin{figure}
\centering
\includegraphics[width=9cm]{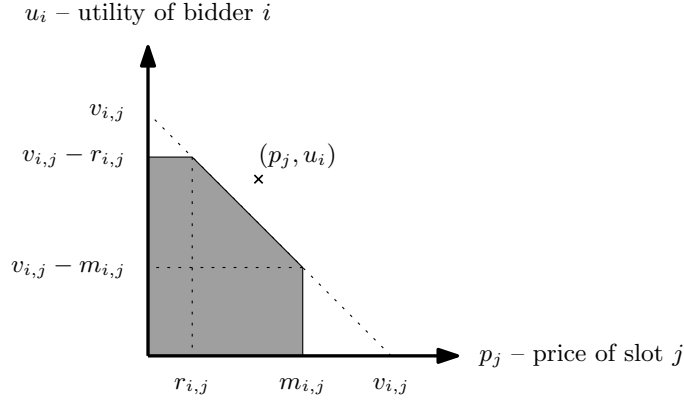}
\caption{Matching is stable whenever for each bidder $i \in I$ and
each slot $j \in J$ the point with coordinates $(p_j, u_i)$ lies
outside the gray region.}
\label{figure:stable}
\end{figure}
Geometric interpretation of inequalities
(\ref{equation:stability1}), (\ref{equation:stability2}),
(\ref{equation:stability3}) is explained in
Figure~\ref{figure:stable}.  Note that if a bidder $i$ is not
interested in a slot $j$, then (\ref{equation:stability2}) is
trivially satisfied.

A feasible matching does not have to be stable, and a stable matching
does not have to be feasible.  However, we will be interested in
matchings that are both stable and feasible, and in addtion bidder-optimal.

\begin{definition}[Bidder Optimality]
\label{def:bidder-optimal}
A stable, feasible matching $(u^*,p^*,u^*)$ is \emph{bidder-optimal} 
if for every stable feasible matching $(u,p,\mu)$ and every bidder $i\in I$
we have $u^*_i \ge u_i$.
\end{definition}

\paragraph{Bidder Preferences.}
To study strategic behavior of bidders in an auction, we need to
model bidder's preferences.
We assume that each bidder is
indifferent among various outcomes as long as her assigned slot (if
any) and payment is the same.  Let us define the utility (payoff) of a bidder
$i$ who is offered a slot $j$ at price $p$ as follows. If $p\le m_{i,j}$,
we set $u = v_{ij} - p$.  If $p>m_{i,j}$, we set $u = -1$.
This payoff, interpreted as a function of the price, is not 
continuous at $p=m_{i,j}$.  If the
bidder is unmatched (at zero price), her payoff is 0. Given a
choice between slot $j_1$ at price $q_1 \le m_{i,j_1}$ and slot $j_2$
at price $p_2 \le m_{i,j_2}$, the bidder prefers the offer with higher
payoff, and is indifferent among offers that have the same payoff.
In particular, the bidder prefers to be not matched to being matched
to a slot $j$ at price that exceeds her maximum price $m_{ij}$.  The
bidder is indifferent between being matched with payoff 0 and not
being matched. 

We call a bidder whose preferences can be described by a 
vector of maximum prices and values a \emph{max-value} bidder.
We point out two classes of bidders that are of interest. 

A \emph{profit maximizing} bidder $i$ only cares about the values $v_{ij}$
he can gain from each position, and seeks to maximize value of the item 
received minus payment. For such bidder we can render the maximum price 
$m_{ij}$ ineffective by setting it to $v_{ij}$.

A \emph{maximum price} bidder is parametrized by a maximum price $m_i$ 
he is willing to pay. He seeks to get the lowest-index position whose price
is less than or equal to $m$.

A more detailed discussion of issues like 
bidder types, their relation to auction mechanisms and
differences between charging per impression and per click is deferred 
to Appendix.

\subsection{Our Results}

Every auction instance in our model has a stable matching by the result of 
\cite{FujishigeTamura2007}. We show
that it also has a bidder-optimal matching, and to give an algorithm to find
it.

\begin{theorem}
\label{theorem:bidder-optimal}
If the auction $(v,m,r)$ is in a ``general position'', it has a unique
\emph{bidder-optimal stable matching}. This matching can be found 
in time $O(nk^3)$.
\end{theorem}

We defer the precise definition of general position to
Definition~\ref{definition:general-position}. In essence, any auction $(v,m,r)$
can be brought into general position by arbitrarily small (symbolic)
perturbations.  In practice this assumption is easily removed by using a
consistent tie-breaking rule.

Consider the following mechanism for auctioning off $k$ items to $n$ 
bidders. The auctioneer (seller) sets an arbitrary minimum price $r_{ij}$ for each 
bidder-item pair. It then solicits vectors of maximum prices
$m_i = (m_{i1}, m_{i2}, \dots, m_{ik})$
and values $v_i = (v_{i1}, v_{i2}, \dots, v_{ik})$ from each bidder $i$.
Finally, the auctioneer computes a bidder-optimal stable matching 
$(u^*, p^*, \mu^*)$ for the 
auction instance $(m,v,r)$. It assigns each bidder $i$ the item (if any) 
$j = \mu^*(i)$ and charges him price $p^*_{j}$ (or $0$ if $\mu^*(i)=\emptyset$).
Let us call this mechanism 
the \emph{Stable Matching Mechanism}. 
Our second technical contributionis to show that the Stable matching Mechanism
is truthful for max-value bidders.

\begin{theorem}[Truthfulness]
\label{theorem:truthfullness}
In the Stable Matching Mechanism, it is a (weakly) dominant strategy for 
each bidder $i$ to submit her true vectors $v_i$ and $m_i$, as long as 
$i$'s preferences can be expressed in the max-value model.
\end{theorem}

\section{An Algorithm to Compute a Bidder-Optimal Matching} 
\label{section:algorithm}

We now describe algorithm \StableMatch 
that computes a feasible and stable matching 
for a given auction instance $(v,m,r)$. Later in Section \ref{section:analysis}
we show that the matching is also bidder-optimal, as long as the auction instance
is in a general position (Definition \ref{definition:general-position}).

The \StableMatch algorithm is an extension of the well known Hungarian Method 
\cite{HungarianMethodWikipedia,Kuhn1955} for computing a maximum-weight 
matching in a bipartite graph. The Hungarian Method is a primal-dual algorithm
that starts with an empty matching and repeatedly increases the size of the matching
using a maximum-weight augmenting path. \StableMatch works the same way, 
except that it is designed to handle events correponding to reaching minimum
and maximum prices.

\StableMatch starts with an empty
matching $(u^{(0)}, p^{(0)}, \mu^{(0)})$ which is defined as follows. Utility
of each bidder $i$ is $u^{(0)}_i = B$, where $B$ is a large enough number, such
that 
$B > \max \{ v_{i,j} ~|~ (i,j) \in I\times J\}$.  Price of each slot $j$ is
$p^{(0)}_j = 0$.  There are no matched pairs, i.e.  $\mu^{(0)} = \emptyset$.

In each iteration, \StableMatch finds an augmenting path, and updates
the current matching
$(u^{(t)}, p^{(t)}, \mu^{(t)})$ to the next matching
$(u^{(t+1)}, p^{(t+1)}, \mu^{(t+1)})$.
The algorithm stops when no more updates can be made, and 
outputs the current matching 
$(u^{(T)}, p^{(T)}, \mu^{(T)})$ at the end of the last iteration.
We now describe an iteration in more detail. To do so, we
introduce the concept of an update graph. 

\begin{definition}[Update graph]
\label{definition:update}
Given an auction $(v,m,r)$, the \emph{update graph} for a matching $(u,p,\mu)$
is a directed weighted bipartite multigraph with partite sets $I$ and $J \cup
\{j_0\}$, where $j_0$ is the \emph{dummy} slot.  The update graph consists of
five types of edges.  For each bidder $i$ and each slot $j \in J$ there is
\begin{itemize}
\itemsep-3pt
\item a \emph{forward edge} from $i$ to $j$ with weight $u_i+p_j-v_{i,j}$, if $p_j \in [r_{i,j}, m_{i,j})$;
\item a \emph{backward edge} from $j$ to $i$ with weight $v_{i,j}-u_i-p_j$, if $(i,j) \in \mu$,
\item a \emph{reserve-price edge} from $i$ to $j$ with weight $u_i + r_{i,j} - v_{i,j}$,
if $u_i + r_{i,j} > v_{i,j}$ and $m_{i,j} > r_{i,j}$,
\item a \emph{maximum-price edge} from $i$ to $j$ with weight $u_i + m_{i,j} - v_{i,j}$,
if $u_i + m_{i,j} > v_{i,j}$ and $m_{i,j} > r_{i,j}$,
\item a \emph{terminal edge} from $i$ to $j_0$ with weight $u_i$ if $u_i > 0$.
\end{itemize}
\end{definition}

An \emph{alternating path} in the update graph starts with an
unmatched bidder vertex $i_0$ with $u_{i_0}>0$, follows a sequence of
forward and backward edges, and ends with a reserve-price, maximum-price or
terminal edge. We place the restriction that all vertices of the
alternating path must be distinct, with the possible exception that
the last vertex is allowed to appear once again along the path.  The
weight $w(P)$ of an alternating path $P$ is the sum of weights of its
edges.

Let $(u^{(t)}, p^{(t)}, \mu^{(t)})$ be a matching and $G^{(t)}$ be the
corresponding update graph. A single iteration of the \StableMatch algorithm
consists of the following steps.
\begin{enumerate}
\item If there is no alternating path, stop and output the current matching.
Otherwise, let $P$ be an alternating path in $G^{(t)}$ of minimum weight.
Let $w^{(t)}(P)$ denote its weight, and let 
$$
P = (i_0, j_1, i_1, j_2, i_2, \dots, j_\ell, i_\ell, j_{\ell+1})  
  \qquad \text{for some } \ell \ge 0 \; .
$$

\item Let $d^{(t)}(i_0, y)$ be the length of the shortest path in $G^{(t)}$
from $i_0$ to any vertex $y$, using only forward and backward edges. 
If a vertex $y$ is not reachable from $i_0$, $d^{(t)}(i_0, y) = \infty$.

\item Compute utility updates for each bidder $i\in I$.
The vector $u^{(t+1)}$ gives the final utilities for the iteration. 
\begin{equation}
\label{equation:utility-update}
u^{(t+1)}_i = u^{(t)}_i - \max\left(w^{(t)}(P) - d^{(t)}(i_0, i),\ 0 \right)
\end{equation}

\item Compute price updates for each slot $j\in J$. 
\begin{equation}
\label{equation:price-update}
p^{(t+)}_j = p^{(t)}_j + \max\left(w^{(t)}(P) - d^{(t)}(i_0, j),\ 0 \right)
\end{equation}
The final prices $p^{(t+1)}_j$ are equal to $p^{(t+)}_j$ with one
exception. 
In case the last edge of $P$ is a reserve-price edge, we set the price
of slot $j_{\ell+1}$, the last vertex of $P$ to be
$p^{(t+1)} = \max(p^{(t+)}, r_{i_{\ell}, j_{\ell+1}})$.

\item Update the assignment $\mu^{(t)}$ along the alternating path $P$
  to obtain the new assignment $\mu^{(t+1)}$. 
\end{enumerate}

We have not specified how should the set of assignment edges be
updated. Before we do that, let us state two invariants maintained by 
\StableMatch. 

\begin{itemize}
\item[(A1)] The matching $(u^{(t)}, p^{(t)}, \mu^{(t)})$ is stable for
  the auction $(v,m,r)$.
\item[(A2)] For every matched pair $(i,j) \in \mu^{(t)}$, $u^{(t)}_i$
  and $p^{(t)}_j$ satisfy (\ref{equation:feasibility1}) and
  (\ref{equation:feasibility2}).
\end{itemize}

An important consequence of invariant (A1) is that forward edges have
non-negative weight. Indeed, it can be easily checked that a forward edge with
a negative weight would be blocking pair. Invariant (A2) guarantees that
backward edges have zero weight. Similarly, invariant (A2) implies that the
weight of every backward edge must be zero.  Finally, each reserve-price,
maximum-price and terminal edges has non-negative weight by definition.

\begin{lemma}
All edge weights in each update graph $G^{(t)}$ are non-negative.
\end{lemma}

With non-negative edge weights, single-source shortest paths can be
computed using Dijkstra's algorithm in time proportional to the square
of the number of vertices reachable from the source. Since no
unmatched vertex is reachable from any other vertex, there are at most
$2k$ reachable vertices at any time, 
thus the shortest alternating path $P$ and
distances $d^{(t)}(i_0,y)$ can be computed in time $O(k^2)$.

Finally, let us deal with updating the assignment $\mu$. Since the
alternating path alternates between using forward (i.e. non-matching)
and backward (i.e. matching) edges, a natural move is to remove all
the matching edges of $P$ and replace them by non-matching edges of
$P$. Care must be taken however to take into account the special
nature of the last edge of $P$ as well as the fact that the last
vertex of $P$ may be visited twice. We consider three cases:

\emph{Case 1:} $P$ ends with a terminal edge, i.e. $j_{\ell+1}$ is the dummy
slot. Flip matching and non-matching edges along the whole length of $P$.
Bidder $i_\ell$ ends up being unmatched, and for $x=0,1,\dots,\ell-1$,
bidder $i_x$ will be matched to slot $j_{x+1}$.

\emph{Case 2:} $P$ ends with a maximum-price edge. Consider two subcases:

\begin{itemize}

\item[(a)] $j_{\ell+1}=j_{\ell}$. This means that the price bidder
  $i_\ell$ was matched to reached his maximum price. Flip matching an
  non-matching edges along $P$. This leaves bidder $i_\ell$ unmatched,
  and for $x=0,1,\dots,\ell-1$ bidder $i_x$ is matched with slot 
  $i_{x+1}$.

\item[(b)] Otherwise, the maximum price was reached on a non-matching
  edge. Keep the matching unchanged. That is,
  $\mu^{(t+1)}=\mu^{(t)}$.

\end{itemize}

\emph{Case 3:} $P$ ends with a reserve-price edge. This is the most
complex case. Consider three subcases:

\begin{itemize}
\item[(a)] Item $j_{\ell+1}$ is unmatched in $\mu^{(t)}$. This case
  increases the size of the matching. For $x=0,1,\dots,\ell$,
  match bidder $i_x$ with slot $j_{x+1}$ .

\item[(b)] Item $j_{\ell+1}$ is matched in $\mu^{(t)}$ and 
  the reserve price $r_{i_\ell,j_{\ell+1}}$ offered by bidder $i_\ell$
  does not exceed the current price $p^{(t+)}_{j_{\ell+1}}$ of the
  slots. Keep the matching unchanged, that is, $\mu^{(t+1)}=\mu^{(t)}$.

\item[(c)] Item $j_{\ell+1}$ is matched in $\mu^{(t)}$ to some 
  bidder $i_{\ell+1}$ and $r_{i_\ell,j_{\ell+1}} > p^{(t+)}_{j_{\ell+1}}$.
  If $P$ is a path, that is, if $P$ does not visit slots $j_{i_\ell}$
  twice, we simply unmatch bidder $i_{\ell+1}$, and flip matching
  and non-matching edges of $P$. (This keeps the size of the matching
  the same, as bidder $i_0$ gets matched and bidder $i_{\ell+1}$
  unmatched.)

  If $P$ visits $j_{\ell+1}$ twice, it must be that $j_{\ell+1}=j_d$
  for some $d$. Note that it is not the case that $d=\ell$, since this
  would mean that $i_\ell$ was matched to $j_{\ell+1}$. This is
  impossible because the reserve price on this edge has been reached
  just now.  This way, the end of $P$ forms a cycle with at least 2
  bidders and 2 slots. We flip the matching and non-matching edges
  along the cycle, but leave the rest of $P$ untouched. 
  This leaves bidder $i_x$ matched to slot $j_{x+1}$, for
  $x=d,d+1,\dots,\ell$.
\end{itemize}

\section{Analysis}
\label{section:analysis}

In this section we show that the \StableMatch algorithm from Section
\ref{section:algorithm} computes a bidder-optimal stable
matching for any auction instance $(v,m,r)$ in general position.

Invariants (A1) and (A2) claimed in the previous section
are enough to show that the resulting matching is feasible and stable.
We prove these invariants and establish a few new ones in 
Appendix \ref{subsection:invariants}.

\begin{lemma}
\label{lemma:feasible-stable}
The matching $(u^{(T)}, p^{(T)}, \mu^{(T)})$ computed by the 
\StableMatch algorithm is feasible and stable.
\end{lemma}

\begin{proof}
Stability follows directly from invariant (A1). Feasibility follows from 
invariant (A2) and the fact that since there are no alternating paths,
it must be that $u^{(T)}_i=0$ for every unmatched bidder $i$.
\end{proof}

\paragraph{Running Time.}
The number of iterations is bounded by $O(nk)$ in 
Lemma \ref{lemma:running-time} below 
(see proof in Appendix \ref{subsection:running-time-proof}). Since each
iteration can be implemented in time $O(k^2)$, this gives us overall running 
time $O(nk^3)$.
\begin{lemma}
\label{lemma:running-time}
\StableMatch finishes after at most $n(2k+1)$ iterations.
\end{lemma}

\subsection{Bidder Optimality}
\label{section:bidder-optimality}

While the matching returned by \StableMatch is always stable and feasible, 
it may not be bidder-optimal. As the following example shows, a 
bidder-optimal matching does not always exist.

\begin{example}
\label{example:no-bidder-optimal-matching}
Consider the case of a single slot and two bidders with
identical maximum bids. There are two stable matchings. In each matching, the
slot is allocated to one of the bidders at maximum price. Each
matching is preferred by one bidder over the other, hence there is no
matching preferred by both of them. 
\end{example}

This example is degenerate in that the maximum bids of both bidders are 
the same. However it turns out that except for such degenerate 
cases, a bidder-optimal matching always exists and \StableMatch will find it.
We make this precise in the following two definitions.

\begin{definition}[Auction graph]
The \emph{auction graph} of an auction $(v,m,r)$ is
a directed weighted bipartite multigraph with partite
sets $I$ and $J \cup \{j_0\}$, where $j_0$ is the \emph{dummy} slot.
The auction graph contains five types of edges. For each bidder $i$ and each slot $j \in J$ there exist
\begin{itemize}
\itemsep-3pt
\item a \emph{forward edge}       from $i$ to $j$ with weight $-v_{i,j}$,  
\item a \emph{backward edge}      from $j$ to $i$ with weight $v_{i,j}$,  
\item a \emph{reserve-price edge} from $i$ to $j$ with weight $r_{i,j} - v_{i,j}$, 
\item a \emph{maximum-price edge} from $i$ to $j$ with weight $m_{i,j} - v_{i,j}$,
\item a \emph{terminal edge}      from $i$ to $j_0$ with weight $0$. 
\end{itemize}
\end{definition}

\begin{definition}[General Position]
\label{definition:general-position}
An auction $(v,m,r)$ is in \emph{general position} if for every bidder
$i$, no two alternating walks in the auction graph that start at bidder
$i$, follow alternating forward and backward edges and end with a
distinct edge that is either a reserve-price, maximum-price or
terminal edge, have the same weight.
\end{definition}

Any auction $(v,m,r)$ can be brought into general position by a
symbolic perturbation. In the algorithm implementation, this can be
achieved by breaking ties lexicographically by the identity of
the final edge of the walk. 

All we need now to prove Theorem \ref{theorem:bidder-optimal} is the following 
lemma, proof of which appears in 
Appendix \ref{subsection:bidder-optimality-proofs}.

\begin{lemma}
\label{lemma:optimality}
Let $(v,m,r)$ be an auction in general position, and let 
$(u',p',\mu')$ be any feasible stable matching. 
Then in any iteration $t$ of \StableMatch, we have that 
$u'_i\le u^{(t)}_i$ for all $i\in I$ and 
$p'_j \ge p^{(t)}_j$ for all $j\in J$.
\end{lemma}

\begin{proof}[Proof of Theorem \ref{theorem:bidder-optimal}]
Consider an auction instance $(m,v,r)$ in general position.
The \StableMatch algorithm on this instance outputs a matching 
$u^*,p^*,\mu^*$ that
is stable and feasible by Lemma \ref{lemma:feasible-stable}. 
Applying Lemma \ref{lemma:optimality} to the current matching after 
the last iteration of the algorithm implies that
$u^*,p^*,\mu^*$ is weakly preferred to any stable matching by every bidder 
and hence is bidder-optimal. Running time of the algorithm follows from 
Lemma \ref{lemma:running-time}.
\end{proof}

\section{Incentive Compatibility}
\label{section:truthfulness}

In this section we will prove Theorem~\ref{theorem:truthfullness}.  A
mechanism based on computing men-optimal stable matching has been
shown to be truth-revealing in several contexts. For the basic stable
matching problem without payments, a concise proof can be found in
\cite{NisanRoughgardenTardosVazirani2007}. For the case of continuous
utilities, a proof was given in \cite{DemangeGale1985}.  Our proof for
the max-value model mimics the overall structure of its
predecessors. First, we show that there is no feasible matching in
which every single bidder would be better off than in the
bidder-optimal matching. (Note that if an agent or set of agents were
to successfully lie about their preferences, the mechanism would still
output a matching that is feasible with respect to the true
preferences.)  This property is known as weak Pareto optimality of the
bidder-optimal matching.

\begin{lemma}[Pareto optimality]
\label{lemma:pareto}
Let $(v,m,r)$ be an auction in general position and let 
$(u^*, p^*, \mu^*)$ be the bidder-optimal matching. 
Then for any matching $(u,p,\mu)$ 
that is feasible for $(v,m,r)$, there is at least one bidder 
$i \in I$ such that $u_i \le u^*_i$.
\end{lemma}

Second, we show that every feasible matching is either stable, or has
a blocking bidder-slot pair that involves a bidder who is not better
off in this matching than in the bidder-optimal matching. 
Versions of the following lemma appear in 
\cite{GaleSotomayor1985, DemangeGaleSotomayor1987, RothSotomayor1990}.
The original statement in a model without money is attributed to J. S. Hwang. 

\begin{lemma}[Hwang's lemma]
\label{lemma:hwang}
Let $(u,p,\mu)$ be a matching that is feasible for an auction $(v,m,r)$ in
general position and let 
$(u^*, p^*, \mu^*)$ 
be the bidder-optimal matching for that auction. Let
$$
I^+ = \{ i \in I \ | \ u_i > u^*_i \} \; .
$$
If $I^+$ is non-empty, 
then there exists a blocking pair $(i,j) \in (I - I^+) \times J$.
\end{lemma}

Proofs of Lemmas \ref{lemma:pareto} and \ref{lemma:hwang} appear in 
Appendix \ref{section:incentive-compatibility-proofs}.
Theorem~\ref{theorem:truthfullness} directly follows from Lemma
\ref{lemma:hwang}. In fact, the lemma implies the following 
stronger statement. 

\begin{theorem}
There is no way for a bidder or a coalition of bidders to manipulate
their bids in a way such that every bidder in the coalition would
strictly benefit from the manipulation.
\end{theorem}

\begin{proof}
Suppose there is a coalition $I^+$ of bidders that can benefit from
submitting false bids. Let $(v,m,r)$ be an auction that reflects the
true preferences of all bidders, and let $(v',m',r)$ be an auction
that reflects the falsified bids. Note that $v'_i = v_i$ and
$m'_i=m_i$ except for bidders $i\in I^+$. 

Let $(u,p,\mu)$ be the bidder-optimal stable matching for the auction
$(v',m',r)$.  First observe that the matching $(u,p,\mu)$ must be feasible for
the true auction $(v,m,r)$. This is because for each bidder $i\in I-I^+$, the
feasibility constraints are the same in both auctions. For bidders $i\in I^+$,
we need to verify that $p_j\le m_{i,j}$ whenever $(i,j)\in \mu$. This follows
because the true bidder-optimal matching $(u^*, p^*, \mu^*)$ respects maximum
prices, and any outcome that respects maximum prices is preferred over an
outcome that doesn't.

Since $(u,p,\mu)$ is feasible, we can apply Lemma \ref{lemma:hwang}
and conclude that there is a pair $(i,j)$ with $i\in I-I^+$ that is
blocking for the auction $(v,m,r)$.
\end{proof}

\section{Conclusions}
\label{section:conclusions}

We have successfully applied the theory of stable matchings to sponsored search 
auctions. Several open questions remain.

\medskip
Fujishige and Tamura \cite{FujishigeTamura2007} propose a general 
model in which a worker can engage several firms and vice versa, of which
ours is a special case. It would be interesting to see if (and under what 
conditions) worker and firm-optimal equilibria exist, and 
whether our strategyproofness result carries through to this very general 
model.

\medskip
Our max-value model assumes a constant ``exchange rate'' in that each
dollar paid by the bidder is perceived as a dollar received by the seller, 
independent of the identity of the bidder and the item.
Suppose the payment is conditioned on some event (such as a user 
clicking or making a purchase), as is common practice. 
At a mutually agreed (say) cost per click, the total revenue 
estimated by the seller may not be equal to the total cost estimated by the 
buyer, if they have different estimates of the probability of a click
occurring. This discrepancy suggests that we introduce an exchange rate
into equations (\ref{equation:feasibility2}) and 
(\ref{equation:stability3}). In such a model with exchange rates,
we do not know if a stable (let alone bidder optimal) matching exists, or
how to find such matching efficiently.

\medskip
Existence of bidder-optimal matchings in our model has clear
implications on the existence of Nash equilibria in (say) GSP auctions
under various assumptions on bidder valuations / preferences.
(For example, can the result of \cite{EvenDarMansour2008} be re-derived and 
extended by using guaranteed existence of bidder-optimal matchings?)

\paragraph{Acknowledgments:}
We would like to thank Hal Varian, Adam Juda and anonymous referees
for helpful comments and pointers to literature.

\smallskip
\nocite{Varian2008}

\bibliographystyle{plain}
\bibliography{stable-matching-auctions}

\newpage

\appendix

\section{Analysis of \StableMatch}
\label{section:bidder-optimality-proofs}
Proofs of statements from Section \ref{section:bidder-optimality}.

\subsection{Invariants}
\label{subsection:invariants}
We establish several invariants that hold throughout the execution of 
the \StableMatch algorithm. These will be used as ingredients in the 
proof of Theorems .
Besides invariants (A1) and (A2) introduced in Section \ref{section:algorithm},
we claim three more invariants.
\begin{itemize}
\item[(A3)] Each unmatched slot has zero price.
\item[(B1)] if a bidder
$i$ is interested in slot $j$ and $u_i^{(t)} + m_{i,j} = v_{i,j}$, then $(i,j)
\not \in \mu^{(t)}$.
\item[(B2)] If a bidder $i$ is interested in a slot $j$ and
$u^{(t)}_i + r_{i,j} = v_{i,j}$, then $(i,j) \in \mu^{(t)}$ or $p^{(t)}_j \ge
r_{i,j}$.
\end{itemize}

All the five invariants are proved by induction on $t$.
Invariants (B1) and (B2) are technical and we omit
their proofs in this version of the paper. 
However, we use them in the induction step to prove the first three
invariants. Both (B1) and (B2) rely on the general position assumption.

\begin{proof}[Proof of the invariants]
The base case, $t=0$, is readily verified. Invariant (A1) follows from that
$u^{(0)}_i = B$ for all $i \in I$, $p^{(0)}_j = 0$ for all $j \in J$, and
hence~(\ref{equation:stability1}) is satisfied. Invariants (A2) and (A3) hold
trivially.

Let us prove that $(u^{(t+1)}, p^{(t+1)}, \mu^{(t+1)})$ satisfies (A3).  Note
that $p^{(t+1)} \ge p^{(t)}$. The slots matched in $\mu^{(t)}$ remain matched
in $\mu^{(t+1)}$, at most one additional slot is matched in $\mu^{(t+1)}$.  The
remaining slots are not reachable from $i_0$ in $G^{(t)}$, since for any such
slot $j$, $p^{(t)}_j = 0$ and for any $i \in I$, $r_{i,j} > 0$ by the general
position assumption, thus there is no forward edge to $j$.  Hence the price of
any such slot $j$ remains zero.

Let us prove that $(u^{(t+1)}, p^{(t+1)}, \mu^{(t+1)})$ satisfies (A1). 
We consider three cases for any pair $(i,j) \in I \times J$:

\emph{Case 1:} $p^{(t)}_j \in [r_{i,j}, m_{i,j})$.  
$(u^{(t)}, p^{(t)},
\mu^{(t)})$ is stable by the induction hypothesis 
and hence $u^{(t)}_i + p^{(t)}_j \ge v_{i,j}$. If $d^{(t)}(i_0, i)
\ge w^{(t)}(P)$, then $u^{(t+1)}_i = u^{(t)}_i$ and $p^{(t+1)}_j \ge
p^{(t)}_j$, thus $u^{(t+1)}_i$ and $p^{(t+1)}_i$
satisfy~(\ref{equation:stability1}).  

On the other hand, if $d^{(t)}(i_0, i) < w^{(t)}(P)$, then
\begin{align}
\label{equation:A1-case1-a}
u^{(t+1)} & = u^{(t)}_i - (w^{(t)}(P) - d^{(t)}(i_0, i)) \; ,
\\
\label{equation:A1-case1-b}
p^{(t+1)}_j & \ge p^{(t+)} \ge  p^{(t)}_j + (w^{(t)}(P) - d^{(t)}(i_0, j)) \; .
\end{align}
Since from $i$ to $j$ there is a forward edge
in $G^{(t)}$, 
\begin{equation}
\label{equation:A1-case1-c}
d^{(t)}(i_0, j) \le d^{(t)}(i_0, i) + (u^{(t)}_i + p^{(t)}_j - v_{i,j}) \; .
\end{equation}
We add~(\ref{equation:A1-case1-a}) to (\ref{equation:A1-case1-b}), subtract (\ref{equation:A1-case1-c}),
and we get that $u^{(t+1)}_i$ and $p^{(t+1)}_j$ satisfy~(\ref{equation:stability1}).

\emph{Case 2:} $p^{(t)}_j \ge m_{i,j}$. Since $p^{(t+1)}_j \ge p^{(t)}_j$,
(\ref{equation:stability2}) holds for $p^{(t+1)}_j$. (This case applies also if
$i$ is not interested in $j$.)

\emph{Case 3:} $p^{(t)}_j < r_{i,j}$ and $i$ is interested in $j$. 
$(u^{(t)}, p^{(t)}, \mu^{(t)})$ is stable by the induction hypothesis
and hence $u^{(t)}_i$ satisfies~(\ref{equation:stability3}).
If $d^{(t)}(i_0, i) \ge w^{(t)}(P)$, then $u^{(t+1)}_i = u^{(t)}_i$ and hence
$u^{(t+1)}_i$ also satisfies~(\ref{equation:stability3}).  

On the other hand, if $d^{(t)}(i_0, i) < w^{(t)}(P)$, then 
\begin{equation}
\label{equation:A1-case3-a}
u^{(t+1)}_i = u^{(t)}_i - (w^{(t)}(P) - d^{(t)}(i_0, i)) \; .
\end{equation}
We claim that in $G^{(t)}$ there is reserve-price edge from $i$ to $j$
and thus 
\begin{equation}
\label{equation:A1-case3-b}
w^{(t)}(P) \le d^{(t)}(i_0, i) + (u^{(t)}_i + r_{i,j} - v_{i,j}) \; .
\end{equation}
To prove the existence of the reserve-price edge we show that $u^{(t)}_i +
r_{i,j} > v_{i,j}$. The non-strict inequality holds since $u^{(t)}_i$
satisfies~(\ref{equation:stability3}).  The strictness follows since,
by the induction hypothesis, $(u^{(t)}, p^{(t)}, \mu^{(t)})$ satisfies (A2) and (B2) .

By subtracting~(\ref{equation:A1-case3-b}) from~(\ref{equation:A1-case3-a}) 
we get that $u^{(t+1)}$ satisfies~(\ref{equation:stability3}).

\medskip

First, let us prove that $(u^{(t+1)}, p^{(t+)}, \mu^{(t)})$ satisfies (A2). Consider any pair
$(i,j) \in \mu^{(t)}$. In $G^{(t)}$ there is a backward edge from $j$ to $i$.
By induction hypothesis, $(u^{(t)}, p^{(t)}, \mu^{(t)})$ satisfies (A2) and hence the backward edge
has zero weight. Hence
\begin{equation}
\label{equation:A2-a}
d^{(t)}(i_0, i) = d^{(t)}(i_0, j) \; .
\end{equation}
Therefore, from the updates~(\ref{equation:utility-update}),
(\ref{equation:price-update}) follows $u_i^{(t+1)} + p^{(t+)}_j = u^{(t)}_i +
p^{(t)}_j$ and hence (\ref{equation:feasibility1}) remains to hold. 

If $w^{(t)}(P) \le d^{(t)}(i_0, i)$, then $p^{(t+)}_j = p^{(t)}_j$ and thus
(\ref{equation:feasibility2}) remains satisfied by $p^{(t+)}_j$. On the other
hand, if $w^{(t)}(P) > d^{(t)}(i_0, i)$, then by the update~(\ref{equation:price-update}) for prices
\begin{equation}
\label{equation:A2-c}
p^{(t+)}_j = p^{(t)}_j + (w^{(t)}(P) - d^{(t)}(i_0,j)) \; .
\end{equation}
We also claim that there exists maximum-price edge from $i$ to $j$ and thus
\begin{equation}
\label{equation:A2-b}
w^{(t)}(P) \le d^{(t)}(i_0, i) + (u^{(t)}_i + m_{i,j} - v_{i,j}) \; .
\end{equation}
To prove the existence of the maximum-price edge we show that $u^{(t)}_i +
m_{i,j} > v_{i,j}$.  The non-strict inequality holds since $p^{(t)}_j \le
m_{i,j}$ and thus $u^{(t)}_i + m_{i,j} \ge u^{(t)}_i + p^{(t)}_j = v_{i,j}$
since by the induction hypothesis $(u^{(t)}, p^{(t)}, \mu^{(t)})$ satisfies
(A2).  Strictness follows since, by the induction hypothesis, $(u^{(t)},
p^{(t)}, \mu^{(t)})$ satisfies (B1).

Summing~(\ref{equation:A2-a}), (\ref{equation:A2-b}),
(\ref{equation:A2-c}) and canceling common terms gives
$p^{(t+)} \le (u^{(t)}_i + p^{(t)}_j - v_{i,j}) + m_{i,j} = m_{i,j}$, where
$u^{(t)}_i + p^{(t)}_j - v_{i,j} = 0$ follows from the induction hypothesis.
Hence, since $p^{(t+)} \ge p^{(t)} \ge r_{i,j}$, (\ref{equation:feasibility2}) remains
to hold for $p^{(t+)}_j$.

Finally, let us prove that $(u^{(t+1)}, p^{(t+1)}, \mu^{(t+1)})$ satisfies
(A2).  For any pair $(i,j) \in \mu^{(t)} \cap \mu^{(t+1)}$ we have already done
it, since $p^{(t+1)}_j = p^{(t+)}_j$.  It remains to consider pairs in
$\mu^{(t+1)} \setminus \mu^{(t)}$. Let $P=(i_0, j_1, i_1, \dots, j_{\ell},
i_{\ell}, j_{\ell+1})$ be the alternating path used to obtain $\mu^{(t+1)}$
from $\mu^{(t)}$.  Any pair $(i,j) \in \mu^{(t+1)} \setminus \mu^{(t)}$ is an
edge lying $P$ and has the form $(i,j) = (i_x, j_{x+1})$. We consider two cases.

\emph{Case 1:} $x < \ell$. In this case $(i,j) = (i_x, j_{x+1})$ is a forward
edge and has weight $u^{(t)}_i + p^{(t)}_j - v_{i, j}$, and
since it lies on a minimum-weight path, 
\begin{equation}
\label{equation:A2-d}
d^{(t)}(i_0, j) = d^{(t)}(i_0, i) + (u^{(t)}_i + p^{(t)}_j - v_{i, j}) \; .
\end{equation}
Since $w^{(t)}(P) \ge d^{(t)}(i_0,i)$ and $w^{(t)}(P) \ge d^{(t)}(i_0,j)$, the updated quantities are
\begin{align}
\label{equation:A2-update-utility}
u^{(t+1)}_i &= u^{(t)}_i - (w^{(t)}(P) - d^{(t)}(i_0, i)) \; ,
\\
\label{equation:A2-update-price}
p^{(t+1)}_j &= p^{(t)}_j + (w^{(t)}(P) - d^{(t)}(i_0, j)) \; .
\end{align}
The equality (\ref{equation:feasibility1}) for $u^{(t+1)}_i$ and  $p^{(t+1)}_j$
follows by summing (\ref{equation:A2-update-utility}),
(\ref{equation:A2-update-price}) and subtracting (\ref{equation:A2-d}). 

Let us verify that $p^{(t+1)}_j$ satisfies (\ref{equation:feasibility2}). Since
$(i, j)$ is a forward edge, $p^{(t)}_j \in [r_{i,j}, m_{i,j})$. By the
induction hypothesis $(u^{(t)}, p^{(t)}, \mu^{(t)})$ is stable, thus $u^{(t)}_i
+ p^{(t)}_j \ge v_{i, j}$, hence $u^{(t)}_i + m_{i,j} > v_{i,j}$ and
consequently in $G^{(t)}$ there is a maximum-price edge from $i$ to $j$ of
weight $u^{(t)}_i + m_{i,j} - v_{i,j}$.  Therefore 
\begin{equation}
\label{equation:A2-e}
w^{(t)}(P) \le d^{(t)}(i_0, i) + u^{(t)}_i + m_{i,j} - v_{i,j} \; .
\end{equation} 
We add (\ref{equation:A2-update-price}) to (\ref{equation:A2-e}) and 
from that we subtract (\ref{equation:A2-d}), we cancel common terms and
we have $p^{(t+1)}_j \le m_{i,j}$. The verification of (\ref{equation:feasibility2}) for $p^{(t+1)}_j$ is finished 
by observing that $p^{(t+1)}_j \ge p^{(t)}_j \ge r_{i,j}$.

\emph{Case 2:} $x=\ell$. Since we assume that $(i,j) = (i_\ell,
j_{\ell+1})$ belongs to $\mu^{(t+1)} \setminus \mu^{(t)}$, it can be neither a terminal edge nor a
maximum-price edge, and thus it must be a reserve-price edge
and has weight $u^{(t)}_i + r_{i,j} - v_{i,j}$.
By the same argument $p^{(t+)}_j \le r_{i,j}$, hence $p^{(t+1)} = r_{i,j}$
and clearly satisfies~(\ref{equation:feasibility2}). Observe that
\begin{align*}
u^{(t+1)} & = u^{(t)} - (w^{(t)}(P) - d^{(t)}(i_0, i)) \; ,
\\
w^{(t)}(P) & = d^{(t)}(i_0, i) + (u^{(t)}_i + r_{i,j} - v_{i,j}) \; .
\end{align*}
Subtracting the two equations shows that $u^{(t+1)}_i$ and $p^{(t+1)}_j$ satisfy~(\ref{equation:feasibility1}).
\end{proof}

\subsection{Proof of Lemma \ref{lemma:running-time}}
\label{subsection:running-time-proof}

\begin{proof}[Proof of Lemma \ref{lemma:running-time}]
Consider the number of edges in the update graph. Initially, the 
graph $G^{(0)}$ has at most $nk$ reserve-price, $nk$ maximum-price and $n$ 
terminal edges. We claim that in each iteration, the number of edges in the 
update graph is reduced by one. 
Since \StableMatch must stop when there are no more 
edges left, this bounds the total number of iterations. 

Consider an iteration $t$ of \StableMatch. We claim that 
in the alternating path $P=(i_0, j_1, i_1,\dots, j_\ell, i_\ell, j_{\ell+1})$,
the last edge $(i,j) = (i_\ell, j_{\ell+1})$ 
will not appear in the update graph $G^{(t+1)}$.
This is easily verified by considering three cases:
\begin{itemize}
\item[\emph{Case 1:}] If $(i,j)$ is a terminal edge, then $w^{(t)}(P) = d^{(t)}(i_0, i) +
u^{(t)}_i$ and hence $u^{(t+1)}_i = u^{(t)}_i - (w^{(t)}(P) - d^{(t)}(i_0, i))
= 0$. 
\item[\emph{Case 2:}] If $(i,j)$ is a maximum-price edge, then $w^{(t)}(P) = d^{(t)}(i_0,
i) + (u^{(t)}_i + m_{i,j} - v_{i,j})$ and hence $u^{(t+1)}_i + m_{i,j} =
u^{(t)}_i - (w^{(t)}(P) - d^{(t)}(i_0, i)) + m_{i,j} = v_{i,j}$. 
\item[\emph{Case 3:}] If $(i,j)$ is a reserve-price edge, then $w^{(t)}(P) = d^{(t)}(i_0,
i) + (u^{(t)}_i + r_{i,j} - v_{i,j})$ and hence $u^{(t+1)}_i + r_{i,j} =
u^{(t)}_i - (w^{(t)}(P) - d^{(t)}(i_0, i)) + r_{i,j} = v_{i,j}$.  
\end{itemize}
The utilities never increase and the prices never decrease throughout the
algorithm, thus the edge $(i_\ell, j_{\ell+1})$ does not appear in any update
graph $G^{(t')}$ for any $t' > t$.
\end{proof}

\subsection{Proof of Lemma \ref{lemma:optimality}}
\label{subsection:bidder-optimality-proofs}
Without loss of generality assume that $(u,p,\mu)$ is such that there does \emph{not}
exist a pair $(i,j) \in \mu$ such that $p_j = m_{i,j}$.  If there was such a
pair, then we can decrease prices of some of the items and increase utilities
of some of the bidders such that $p_j < m_{i,j}$. This is possible because of
the general position assumption. See full version of the paper. 

We prove Lemma~\ref{lemma:optimality} by induction on $t$. The base case, $t=0$,
trivially holds true, since by feasibility of $(u',p',\mu')$, $p'_j \ge 0$ for
all $j \in J$ and $u'_i \le B$ for all $i \in I$. In the inductive case, assume
that $u^{(t)} \ge u'$ and $p^{(t)} \le p'$. We first prove that 
\begin{proposition}
\label{proposition:optimality-update-lemma}
$u^{(t+1)} \ge u'$ and $p^{(t+)} \le p'$. 
\end{proposition}

We look ``continuously'' at updates~(\ref{equation:utility-update}) and~(\ref{equation:price-update}).
For that purpose we define for each $i \in I$ a continuous non-increasing function $u_i(x)$,
$$
u_i(x) = u^{(t)}_i - \max\left(x - d^{(t)}(i_0, i),\ 0 \right) \; , 
$$
and for each $j \in J$ a continuous non-decreasing function $p_j(x)$,
$$
p_j(x) = p^{(t)}_j + \max\left(x - d^{(t)}(i_0, j),\ 0 \right) \; .
$$
Clearly, $u^{(t+1)} = u(w^{(t)}(P))$ and $p^{(t+)} = p(w^{(t)}(P))$.
To prove that $u^{(t+1)} \ge u'$ and $p^{(t+)} \le p'$, 
suppose by contraction that there exists $y \in [0, w^{(t)}(P)]$
such that either $u_i(y) < u'_i$ for some $i \in I$ or $p_j(y) > p'_j$ for some $j \in J$.
We choose infimal such $y$. Clearly, $u(y) \ge u'$, $p(y) \le p'$ and $y < w^{(t)}(P)$.
Consider the sets
\begin{align*}
I' &= \{ i \in I ~|~ u_i(y) = u'_i \text{ and } d^{(t)}(i_0, i) \le y \} \; , \\
J' &= \{ i \in J ~|~ p_j(y) = p'_j \text{ and } d^{(t)}(i_0, j) \le y \} \; . 
\end{align*}
\begin{claim}
Each slot $j \in J'$ is matched in $\mu^{(t)}$ to some $i \in I'$.
\end{claim}
\begin{proof}[Proof of the Claim]
Let $j \in J'$. If $j$ was unmatched, then either $d^{(t)}(i_0, j) = w^{(t)}(P)$ or
$d^{(t)}(i_0, j) = \infty$; however both options contradict the choice of $y$ and that $j\in J'$. Thus $j$ is
matched to some $i \in I$, hence in $G^{(t)}$ there is a backward edge from $j$ to
$i$ and thus $d^{(t)}(i_0, i) = d^{(t)}(i_0, j)$ and therefore $u_i(y) + p_j(y)
= v_{i,j}$. Further, invariants (A2) and (B1) imply that $p^{(t)}_j \in [r_{i,j}, m_{i,j})$. 
Consequently, there is a maximum-price edge from $i$ to $j$, $w^{(t)}(P)
\le d^{(t)}(i_0, i) + (u^{(t)}_i + m_{i,j} - v_{i,j})$, and hence $p'_j =
p_j(y) < p^{(t+)}_j = p^{(t)} + (w^{(t)}(P) - d^{(t)}(i_0, j)) \le m_{i,j}$.
Therefore $p'_j \in [r_{i,j}, m_{i,j})$, and since $(u',p',\mu')$ is stable,
$u'_i + p'_j \ge v_{i,j}$ and hence $u_i(y) = v_{i,j} - p_j(y) = v_{i,j} - p'_j
\le u'_i$.  On the other hand, by infimality of $y$, $u_i(y) \ge u'_i$. Thus $i
\in I'$.
\end{proof}

\begin{claim}
Each bidder $i \in I'$ is matched in $\mu'$ to some $j \in J'$.
\end{claim}

\begin{proof}[Proof of the Claim]
Since in $G^{(t)}$ there is a terminal edge from $i$ to the dummy slot, $w^{(t)}(P) \le d^{(t)}(i_0, i) + u^{(t)}_i$.
Hence
\begin{multline*}
u_i' = u_i(y) = u^{(t)}_i - (y - d^{(t)}(i_0, i)) \\ > u^{(t)}_i - (w^{(t)}(P) - d^{(t)}(i_0, i)) \ge 0 \; ,
\end{multline*}
and thus bidder $i$ is matched in $\mu'$ to some slot $j \in J$. 

By feasibility of $(u',p',\mu')$, $p'_j \in [r_{i,j}, m_{i,j}]$. By the
assumption made at the beginning $p_j \neq m_{i,j}$.  Therefore in $G^{(t)}$
there is a forward edge from $i$ to $j$ and thus 
\begin{equation}
\label{equation:claim2-a}
d^{(t)}(i_0, j) \le d^{(t)}(i_0, i) + (u^{(t)}_i + p^{(t)}_j - v_{i,j}) \; .
\end{equation}
Clearly, since $i \in I'$,
\begin{equation}
\label{equation:claim2-b}
u_i(y) = u^{(t)}_i - (y - d^{(t)}(i_0, i)) \; .
\end{equation}
By the price update rule 
\begin{equation}
\label{equation:claim2-c}
p_j(y) \ge p^{(t)}_j + (y - d^{(t)}(i_0, j)) \; .
\end{equation}
We add (\ref{equation:claim2-b}) to (\ref{equation:claim2-c}) and subtract from that (\ref{equation:claim2-a}) and we obtain 
$$
p_j(y) \ge v_{i,j} - u_i(y) \; .
$$
Hence, since by feasibility of $(u',p',\mu')$, $u'_i + p'_j = v_{i,j}$, we have
$$
p_j(y) \ge v_{i,j} - u_i(y) = v_{i,j} - u'_i = p_j' \; .
$$
Recalling that $p(y) \le p'$ we see that $p_j(y) = p'_j$.

Subtracting (\ref{equation:claim2-b}) from (\ref{equation:claim2-a}) and cancelling common terms
we have
$$
d^{(t)}(i_0,j) \le y + (u_i(y) + p^{(t)}_j - v_{i,j}) \; .
$$
We upper-bound the right side of the inequality using that $u_i(y) = u'_i$, $p^{(t)}_j \le p_j(y)$ and $u'_i + p'_j = v_{i,j}$
and we have
$$
d^{(t)}(i_0,j) \le y + (u'_i + p'_j - v_{i,j}) = y \; . 
$$
Thus $j \in J'$.
\end{proof}

From the two claims it follows that $|I'| = |J|'$ and that $\mu^{(t)}$
bijectively matches $I'$ with $J'$. In particular $i_0 \not \in I'$. Choose $j
\in J'$ with smallest $d^{(t)}(i_0, j)$.  Consider the minimum-weight path in
$G^{(t)}$ from $i_0$ to $j$ which uses only forward and backward edges.  The
vertex on the path just before $j$ is a bidder $i \not \in I'$.  Clearly, $y
\ge d^{(t)}(i_0, j) > d^{(t)}(i_0, i)$ and hence $u_i(y) < u'_i$.  There is a
forward edge from $i$ to $j$, thus $p^{(t)}_j \in [r_{i,j}, m_{i,j})$ and also
$u_i(y) + p_j(y) = v_{i,j}$, and hence (*) $u'_i + p'_j < v_{i,j}$. Since in
$G^{(t)}$ there is a maximum-price edge from $i$ to $j$, $p'_j = p_j(y) <
m_{i,j}$, which together with (*) contradicts stability of $(u', p', \mu')$. 
This proves Proposition~\ref{proposition:optimality-update-lemma}.

To prove Lemma~\ref{lemma:optimality} it remains to show that $p^{(t+1)} \le p'$. This amounts to show that if
$(u^{(t+1)}, p^{(t+1)}, \mu^{(t+1)})$ was obtained from $(u^{(t)}, p^{(t)},
\mu^{(t)})$ by updating along an alternating path $P$ of which the last edge,
$(i,j)=(i_{\ell}, j_{l+1})$,  was a reserve-price edge and $p^{(t+)}_j <
r_{i,j}$, then 
\begin{equation}
\label{equation:optimality-a}
r_{i,j} \le p'_j \; .
\end{equation}
Since $(u',p',\mu')$ is stable, either $u'_i
+ p'_j \ge v_{i,j}$ or $p'_j \ge m_{i,j}$.  In former case, (\ref{equation:optimality-a})
follows from that $u^{(t+1)}_i = v_{i,j} - r_{i,j}$,
Proposition~\ref{proposition:optimality-update-lemma} and that $(u',p', \mu')$ is stable.
In latter case, (\ref{equation:optimality-a})
follows since the presence of the reserve-price edge from $i$ to $j$ guarantees
that $m_{i,j} > r_{i,j}$.

\section{Proofs of Incentive Compatibility}
\label{section:incentive-compatibility-proofs}

\begin{proof}[Proof of Lemma \ref{lemma:pareto}]
For the sake of contradiction, suppose that there is a feasible
matching $(u,p,\mu)$ such that $u_i>u^*_i$ for all $i\in I$.  Note
that every bidder must be matched in $\mu$, since $u_i > u^*_i\ge 0$.

For each bidder $i\in I$, consider the slot $j = \mu(i)$ matched to
bidder $i$ in the matching $\mu$. Since the pair $(i,j)$ is not
blocking for the bidder-optimal matching $(u^*,p^*,\mu^*)$, it must be
that $p^*_j > p_j$. In particular, the existence of $\mu$ implies that
there must be $n$ slots with positive prices in the bidder-optimal
matching $\mu^*$, and that these slots are matched in $\mu$ as well.

If a slot ever becomes matched to a bidder in the \StableMatch algorithm,
it will never become unmatched. Thus before the last iteration, at most
$n-1$ slots have positive prices. Suppose the last iteration,
iteration $T-1$, increases the size of the matching to $n$, and let $j$
be the last slot to be matched.  Let $i' = \mu(j)$ be the bidder
matched to $j$ in the hypothetical matching $\mu$.

Let $P$ be the shortest alternating path found in Step 1 of the last
iteration of \StableMatch. Recall that the first vertex
of the path is denoted by $i_0$ and $w^{(T-1)}(P)$ denotes its length.
If $P$ ends with the reserve-price edge $(i,j)$, it must
be that $i$ and $j$ are matched in both $\mu$ and $\mu^*$ at the same
reserve price, contradicting our assumption that $u_i > u_i^*$.

On the other hand, if $P$ does not end with the reserve-price edge
$(i,j)$, we show that there is a shorter alternating path $P'$ that
does include this edge, which again leads to a contradiction.
From Step 3 of the last iteration we have $u_i^{(T-1)} - u_i^* =
w^{(T-1)}(P) - d^{(T-1)}(i_0, i)$.  Let $s$ be the length of the reserve
price edge $(i,j)$; recall from Definition \ref{definition:update}
that $s = u^{(T-1)}_i + r_{i,j} - v_{i,j}$.  Now consider the alternating 
path $P'$ that consists of the shortest path from $i_0$ to $i$ followed by the
reserve price $(i,j)$ edge. We have
$$
w^{(T-1)}(P) - w^{(T-1)}(P') = u_i^{(T-1)} - u_i^* - s = v_{i,j} - r_{i,j} - u^*_i\, .
$$ 
Since $u^*_i < u_i \le v_{i,j}-r_{i,j}$, this difference is positive
and hence $P'$ must be a shorter alternating path than $P$.
\end{proof}

\begin{proof}[Proof of Lemma \ref{lemma:hwang}]
Without loss of generality assume that $(u,p,\mu)$ is such that there does \emph{not} exist a
pair $(i,j) \not \in \mu$ such that $u_i + r_{i,j} = v_{i,j}$.  If there was such a pair,
then we can decrease prices of some of the items and increase utilities of some
of the bidders such that $u_i + r_{i,j} > v_{i,j}$. (This is possible because
of the general position assumption. See full version of the paper.) The set
$I^+$ would only grow by such operation.

Let us denote by $\mu(I^+)$, $\mu^*(I^+)$ the set of slots matched to
bidders in $I^+$ in matching respectively $\mu$, $\mu^*$. We
consider two cases:

\emph{Case 1:} $\mu(I^+) \neq \mu^*(I^+)$. For any $i \in I^+$ we have $u_i > u^*_i \ge 0$
and hence each bidder in $I^+$ is matched in $\mu$ to some slot. There exists a slot $j \in \mu(I^+)$,
$j \not \in \mu^*(I^+)$. Let $i = \mu(j)$. Since $i \in I^+$, $u_i > u^*_i$. 

We argue that $p_j < p^*_j$: By the general position assumption $p^*_j \neq
m_{i,j}$, and hence by feasibility of $(u,p,\mu)$, $p_j \in [r_{i,j}, m_{i,j})$
and $u_i + p_j = v_{i,j}$.  Hence $u^*_i + p^*_j \ge v_{i,j}$. Therefore $p^*_j
\ge v_{i,j} - u^*_i > v_{i,j} - u_i = p_j$.

In particular, $j$ is matched in $\mu^*$ to some $i'$, and by the choice of
$j$, $i' \not \in I^+$. Thus $u_{i'} \le u^*_{i'}$.  By feasibility of $(u^*,
p^*, \mu^*)$, $p^*_j \in [r_{i',j}, m_{i', j}]$ and $u^*_{i'} + p^*_j = v_{i',
j}$. By the assumption on  $(u,p,\mu)$ that we made at the beginning of the
proof, $u_{i'} \neq v_{i',j} - r_{i',j}$.  

Now, it is not hard to see that $(i',j)$ is blocking pair for $\mu$.  This is
because 
\begin{align*}
p_j & < p^*_j \le m_{i,j} \; , 
\\
u_{i'} & \le u^*_{i'} = v_{i',j} - p^*_j \le v_{i',j} - r_{i,j} \; \text{ and } \\
u_{i'} & \neq v_{i',j} - r_{i',j} \; , 
\\
u_{i'} + p_j & < u^*_{i'} + p^*_j = v_{i', j} \; .
\end{align*}

\emph{Case 2:} $\mu(I^+) = \mu^*(I^+) = J^+$. 
Since $u_i > u^*_i$ for $i \in I^+$, by
stability of $(u^*, p^*, \mu^*)$ it follows that $p_j < p^*_j$ for $j \in
J^+$.

Consider a reduced auction $(v',m',r')$ on the set of bidders $I^+$
and set of slots $J^+$. We set the reserve prices to reflect the influence 
of bidders in $I\setminus I^+$. More specifically, 
let $I' = \{i\in I\setminus I^+ ~|~ u^*_{i'} \ge v_{i',j} - r_{i', j}\}$.
For every $i \in I^+$ and $j \in J^+$, we set
$$
r'_{i,j} = \max \big( r_{i,j}, 
             \max_{i'\in I'}  \min (m_{i', j}, v_{i',j}-u^*_i)  \big) \,.
$$
We also set $v'_{i,j} = v_{i,j}$ and $m'_{i,j} = m_{i,j}$ except that if
$m_{i,j} \le r'_{i,j}$ we set $m'_{i,j} = -1$.  It is not hard to show
that if $v,m,r$ is in general position, then so is $(v',m',r')$, using
the fact that each utility $u^*_i$ was at some point set to be equal to the 
length of some alternating walk in the auction graph. 

Now consider the matchings $\mu$ and $\mu^*$ restricted to the sets
$I^+$, $J^+$. If the restricted $\mu$ is not feasible for $(v',m',r')$,
it must be because $p_j < r_{i,j}$ for some position $j = \mu(i)$.
This can only happen if $r'_{i,j} > r_{i,j}$ and hence $r'_{i,j} =
\max(m_{i',j}, v_{i',j} - u^*_{i'})$ for some bidder $i'\in I\setminus
I^+$.

On the other hand, it is easy to check that the restricted matching $\mu^*$ is
feasible, stable and bidder-optimal for the auction $(v',m',r')$.
If the restricted $\mu$ is feasible for this auction, by Lemma 
\ref{lemma:pareto}, there is a bidder $i\in I^*$ such that $u_i \le u^*_i$.
This however contradicts the definition of the set $I^+$.
\end{proof}

\section{Modeling Advertising Auctions}
\label{section:extensions}

In this section, we will present examples of auction mechanisms
commonly used in sponsored search. We will show
how to model these mechanisms in our max-value model. In the 
next section we give examples of novel combined mechanisms
that can be implemented in our model.

\subsection{Existing Mechanisms}

\noindent
{\bf Translating between impressions and clicks.}  Typically, an
auction is run to determine the placement of ads every time a results
page is rendered; however, the advertiser only pays when a user
actually clicks on the ad. It is straightforward to translate between
the pay-per-impression and the pay-per-click model, provided that we
know the probability $\ctr$ that a user will click on the ad:
paying $p_c$ per click is the same in expectation as paying $p_i =
\ctr \cdot p_c$ per impression. In the following, let $\ctr_{i,j}$ be the
probability that a user clicks on ad $i$ if it is displayed in
position $j$ (and that this probability does not vary depending on the
set of competing ads shown on the page). 
The \emph{click separability assumption} says that 
$\ctr_{i,j} = q_i \cdot \alpha_j$ is the product of a quality score $q_i$
of the advertiser and a \emph{position normalizer} $\alpha_j$ specific
to the position $j$. Typically the position normalizers are assumed to
be decreasing, i.e. $\alpha_1\ge \alpha_2\ge\dots\ge\alpha_k$.

\noindent
{\bf GSP pay-per-impression.} In a Generalized Second Price auction, each
advertiser $i$ submits a single number $b_i$ as her bid, which is the
maximum amount she is willing to pay for displaying her ad. The
auctioneer orders bidders in decreasing order of their bids, and
assigns the first $k$ advertisers to the $k$ available slots in this
order. The $i$-th allocated advertiser pays amount equal to the
$(i+1)$-st bid for each impression.

\noindent
{\bf GSP pay-per-click.} An alternative is to charge the advertiser
only in the event of a click on her ad. The bid $b_i$ is interpreted
as a maximum the advertiser is willing to pay for a click. Again, the
advertisers are ordered by their per-click bid, and each allocated
advertiser pays the next highest bid in the event of a click.  In a
quality-weighted variant, the ads are ordered by the product of their
quality score $q_i$ and bid $b_i$; the $i$-th advertiser pays $b_{i+1}
\frac{q_{i+1}}{q_i}$ in the event of a click.  Note that 
the expected cost per impression 
$b_{i+1}\frac{q_{i+1}}{q_i} \ctr_{i,i}$ depends not only on the next
highest bid but also on the position, as long as the probability 
$\ctr_{i,j}$ of clicking on the ad $i$ in position $j$ depends on the position.
Thus, there is no direct way to translate a per-click bid to a per-impression 
bid, without looking at the competitor's bids. 

\noindent
{\bf The VCG mechanism for profit-maximizing bidders.}
In a variant of the VCG mechanism considered e.g. in 
\cite{AggarwalGoelMotwani2006}, each bidder $i$ states her
value $V_i$ for a click. The auctioneer derives the expected 
value of each slot $v_{i,j} = V_i \cdot\ctr_{i,j}$ for that bidder by 
using an estimate $\ctr_{i,j}$ of the probability that 
the ad $i$ would be clicked on if placed in position $j$.
The auctioneer computes a maximum-weight matching in the bipartite 
graph on bidders and positions with $v_{i,j}$ as edge weights. 
The maximum weight matching $\mu^*$ gives the final allocation. 
For pricing, the VCG formula sets the price per impression of slot $j=\mu^*(i)$ to be
$p_j = \sum_{k\in I\setminus\{i\}} v_{k, \mu'(k)} - v_{k, \mu^*(k)}$
where $\mu'$ is a maximum-weight matching with the set of bidders 
$I\setminus\{i\}$. Note that the per-impression price $p_j$ can be translated 
to a per-click price by charging bidder $i$ price $p_j / \ctr_{i,j}$ for
each click. (Similar translation can be done for a generally defined user action
other than a click, as long as the probability of the action can be estimated.)

For each of the above mechanisms, we define a corresponding type of bidder
in the max-value model. 

\noindent
{\bf Max-per-impression bidder} has a target cost per impression
$b_i$. She prefers paying $b_i$ or less per impression to any outcome
where she pays more than $b_i$. Given that her cost per impression is
at most $b_i$, she prefers higher (with lower index) position to lower
position. Given a fixed position, she prefers paying lower price to higher
price.
A max-per-impression bidder $i$ can be translated into the max-value
model by setting her $m_{i,j} = b_i$ for all positions $j \in J$,
and setting her value $v_{i,j} = M(k+1-i)$ where $M$ is a sufficiently
large number ($M > b_i$ is enough).

\noindent
{\bf Max-per-click bidder} differs from a max-per-impression bidder in
that she is not willing to pay more than $b_i$ per click.  We
translate her per-click bid into our framework using predicted click
probabilities: set $m_{i,j} = b_i \cdot \ctr_{i,j}$ for $i \in I$
and $v_{i,j} = M(k+1-i)$ where $M > b_i \max_j \ctr_{i,j}$.

\noindent
{\bf Profit-maximizing bidder} seeks the position and payment that
maximizes her expected profit (value from clicks minus payment).
If we assume that her value per click is $V_i$, 
such bidder is modeled by setting 
$v_{i,j} = m_{i,j} = V_i \cdot \ctr_{i,j}$.

We formalize the correspondence between the mechanisms and corresponding bidder
types in the following theorem.

\begin{theorem}
\label{thm:simulation}
The outcome (allocation and payments) of a 
(1) per-impression GSP, 
(2) per-click GSP, 
(3) VCG auction, respectively
is a bidder-optimal stable matching for a set of 
(1) max-per-impression bidders, 
(2) max-per-click bidders, 
(3) profit-maximizing bidders,
respectively.
\end{theorem}

\begin{proof}
Part (3) of the theorem has first been shown by \cite{Leonard1983}. 
Chapter 7 of \cite{RothSotomayor1990} as
well as \cite{BikhandaniOstroy2006} discuss the relationship of the
VCG mechanism for assignments and stable matchings. 

We give a proof for part (1), per-impression GSP.
The proof of part (2) for per-click GSP is very similar and is omitted.
For simplicity, we assume that $n>k$ and all reserve prices are zero.
Let $b_1 > b_2 >
\dots > b_n$ be the per-impression bids of the bidders. Without loss
of generality, the bidders are ordered by decreasing order of their
bids. (By the general position assumption, assume bids are distinct.)

Recall that we encode a max-per-impression bidder by setting 
$v_{i,j}=M(k-j+1)$ and $m_{i,j}=b_i$. 
The matching produced by the GSP auction is as follows:
the matched pairs are 
$\mu = \{(1,1), (2,2), \dots, (k,k)\}$, 
bidder's utilities
$u_i = M(k-i+1)-b_{i+1}$ 
for $1\le i\le k$, $u_i=0$ for $i>k$, 
and prices $p_i = b_{i+1}$ for $i=1,2,\dots,k$.
It is easy to verify that this matching is feasible and stable
according to Definitions \ref{def:feasibility} and
\ref{def:stability}.

First we show that any feasible matching in which
the assignment is different from $\mu$ is not stable. Indeed, such a
matching $(u',p',\mu')$ must have a bidder $i\le k$ such that $i$ was
not allocated a slot among the first $i$ slots, and a slot $j\le i$
that is either unmatched or matched to some bidder $i'>i$.  

From feasibility we have that $p_j=0$ if slot $j$ is unmatched and
$p_j\le b_{i'}$ in case it is matched. In either case, $p_j <
b_i$. Also, since bidder $i$ is matched to some slot $j'>i$ (or
unmatched), we know that $u'_i \le v_{i,j'} = M(k-j'+1)$.  We now
claim that $(i,j)$ is a blocking pair.  Since $v_{i,j} - u'_i \ge M
[(k-j+1) - (k-j'+1)] \ge M$, inequalities (\ref{equation:stability1})
and (\ref{equation:stability3}) are violated, and since $p'_{j} <
b_i$, inequality (\ref{equation:stability2}) is violated as well.

Now consider any matching with the assignment
$\mu=\{(1,1),\dots,(k,k)\}$. It is easy to verify that in order to be
stable, it must be that $p_{i} \ge b_{i+1}$, otherwise the pair
$(i+1,i)$ would be a blocking pair. Hence the matching with prices
$p_i=b_{i+1}$ has the lowest possible prices and hence is
bidder-optimal.
\end{proof}

\noindent
{\bf Minimum prices.} 
Some search engines impose a minimum price $r_i$ for each 
ad (for example, based on perceived quality of the ad). In GSP, 
only bidders whose bid is above the reserve price can participate. The 
allocation is in decreasing order of bids, and each bidder pays the maximum of
her reserve price and the next bid. 
Minimum GSP prices are easily translated to the max-value model by setting
$r_{ij} = r_i$ (if paying per impression) or $r_{ij} = r_i\cdot \ctr_{i,j}$
(if paying per click). Our model allows for separate reserve prices for 
different slots (e.g. higher reserve price for certain premium slots)
that are not easily implemented in the GSP world.

\subsection{New Auction mechanisms}
\label{section:new-auctions}

Let us give a few examples of new auction mechanisms that are 
special cases of the max-value model.

\noindent
{\bf GSP with arbitrary position preferences.}
Consider an advertiser $i$ who wishes for her ad to appear only in certain 
slots. For example,
\cite{AggarwalMuthukrisnanFeldman2006} propose a GSP variant
in which each bidder has the option to specify a prefix 
of positions $\{1,2,\dots,\beta_i\}$ for some $\beta_i$ 
she is interested in and exclude
the remaining slots. 
Also, tools like Google's Position Preference allow
the advertiser to specify arbitrary position intervals
$[\alpha_i,\beta_i]$. 
We are however not aware of any published work that discusses more
sophisticated position preferences. One would imagine that 
in the world of content advertising where there may be multiple
areas designed for ads on a single page, having a richer language
in which to express the preferences over slots would be beneficial to the 
advertiser. Such preferences are readily expressible in the max-value model.\qed

\noindent
{\bf Combining click and impression bidders in GSP.}
Since both pay per click and pay per impression models are widely used in 
practice, it is useful to have a way of combining these two bidding modes. 
This can be easily done by computing a stable matching for a mixed pool 
of bidders. The following simpler approach is not appropriate, as it does 
not have the proper incentive structure. 

Suppose we allow each bidder $i$ to specify both a maximum price
$b_i$, as well as a payment type $\tau_i\in \{{\cal I}, {\cal C}\}$.
A naive combined auction orders bidders by decreasing $b_i$. Each
advertiser with $\tau_i={\cal I}$ is charged the next highest bid
$b_{i+1}$ for showing the ad. Each advertiser with $\tau_i={\cal C}$
is charged $b_{i+1}$ in the event that the user clicks on the
ad. Note, this scheme gives advertisers a strong incentive to report
$\tau_i={\cal C}$ regardless of their true type (as long as the
probability of user clicking is less than 1).

To offset this incentive, the auctioneer may introduce multipliers
$0<q_{\cal C}<1$ and $q_{\cal I}=1$ and set the effective bid of each
bidder to be $b^{\mathrm{eff}}_i = b_i q_{\tau_i}$.  In the modified
GSP auction, bidders are be sorted by their effective bid. Each bidder
$i$ who reports type $\tau_i=\cal{I}$ is charged
$b^{\mathrm{eff}}_{i+1}$ for each impression, while each bidder
reporting $\tau_i=\cal{C}$ is charged $b^{\mathrm{eff}}_{i+1} /
q_{\cal C}$ in the event of a click.

For any value of $0<q_{\cal C}<1$, there is a simple
instance in which some bidder can gain by misreporting her type.  Let
$\ctr_1$ and $\ctr_2$ be the probability that an user will click on an
ad in position 1 and 2 respectively. Assume this probability is the
same for all ads, and that $\ctr_1 > \ctr_2$.  Suppose that the first
slot is won by a bidder of type $\cal I$, the second slot is won by a
bidder of type $\cal C$, and that there is at least one more bidder
with positive bid.  If $q_{\cal C} > \ctr_2$, the bidder in the second
position can lower her overall cost while keeping the same position by
reporting type $\cal C$ and keeping the same effective bid.  On the
other hand, if $q_{\cal C} < \ctr_1$, bidder in the first position can
lower her cost by reporting type $\cal I$, and adjusting her bid so
that her effective bid stays the same. \qed

\noindent
{\bf Diverse bidders.}  There are many types of bidders with different
goals. Some like to think in terms of a maximum price per click or
impression. Some prefer to target only certain positions (e.g. top of
the page) for consistency or branding reasons. Others try to maximize
their profit and are able to estimate the value of a specific user
action. Each bidder may specify her goal in a language familiar to her.
We are not aware of any prior research on auction 
mechanisms for such diverse set of bidders. \qed

\section{Lattice Property}
\label{section:lattice-property}

The set of feasible and stable outcomes in both the stable marriage and the assignment model
has the algebraic structure of a lattice (see e.g. Chapter 3 in \cite{RothSotomayor1990}).
This result can be carried over to our assignment model with minimum 
and maximum prices as well. 
The following lemma can be proved using ideas and techniques 
from Section \ref{section:bidder-optimality-proofs}. The proof is relatively 
long and tedious and is omitted.

\begin{lemma}[Lattice property]
\label{lemma:lattice-property}
Let $(v,m,r)$ be an auction in general position.
If $(u^A, p^A, \mu^A)$ and $(u^B, p^B, \mu^B)$ are two feasible stable matchings for $(v,m,r)$,
then there exists a feasible stable matching $(u^C, p^C, \mu^C)$ for $(v,m,r)$ such that
\begin{align*}
u^C_i &= \max\{ u^A_i, u^B_i \} & \text{for each $i \in I$,} \\
p^C_j &= \min\{ p^A_j, p^B_j \} & \text{for each $j \in J$,}
\end{align*}
and there exists a feasible stable matching $(u^D, p^D, \mu^D)$ for $(v,m,r)$ such that
\begin{align*}
u^D_i &= \min\{ u^A_i, u^B_i \} & \text{for each $i \in I$,} \\
p^D_j &= \max\{ p^A_j, p^B_j \} & \text{for each $j \in J$.}
\end{align*}
\end{lemma}

The set $\cal M$ of feasible and stable matchings for an auction $(v,m,r)$
is non-empty by \ref{lemma:feasible-stable}. 
If the auction instance is in general position we know that 
$\cal M$ is also a lattice by Lemma \ref{lemma:lattice-property}. 
It is not hard to see that $\cal M$ is closed and bounded, and hence must have
a minimum and maximum element. This gives us an alternate way of proving that
a bidder-optimal stable matching exists.

\end{document}